\documentclass[12pt]{article}
\usepackage{graphicx}

\newcommand \bra[1]{\left< {#1} \,\right\vert}
\newcommand \ket[1]{\left\vert\, {#1} \, \right>}

\newcommand{\bea}{\begin{eqnarray}}
\newcommand{\eea}{\end{eqnarray}}
\newcommand{\simgt}{\hbox{ \raise3pt\hbox to 0pt{$>$}\raise-3pt\hbox{$\sim$} }}
\newcommand{\simlt}{\hbox{ \raise3pt\hbox to 0pt{$<$}\raise-3pt\hbox{$\sim$} }}
\newcommand \vc[1]{{\bf {#1}}}
\newcommand{\clfn}{\setcounter{footnote}{0}}

\begin{document}
\begin{titlepage}
\title{Quarkonium Spectroscopy and Perturbative QCD:\\
A New Perspective\thanks{
Talk given at ``Accelerator and Particle Physics Institute
(APPI 2001)'', Morioka, Japan, Feb.~20--22, 2001.
} \vspace{2cm}}
\author{Y.~Sumino
\\ \\ Department of Physics, Tohoku University\\
Sendai, 980-8578 Japan
}
\date{}
\maketitle
\thispagestyle{empty}
\vspace{-4.5truein}
\begin{flushright}
{\bf TU--623}\\
{\bf May 2001}
\end{flushright}
\vspace{5.0truein}
\begin{abstract}
{\small
We report new aspects of the recent theoretical progress in
heavy quarkonium physics.
(1) Contrary to wide beliefs, 
the gross structure of the bottomonium spectrum is
described well by the non-relativistic boundstate 
theory based on perturbative QCD.
(2) This leads to a new physical picture of the bottomonium states:
the boundstate mass is composed mainly of the 
self-energies of $b$ and $\bar{b}$ accumulated inside the boundstate.
(3) A connection to the conventional phenomenological
potential-model approaches is provided.
}
\end{abstract}
\vfil

\end{titlepage}

\section{Introduction}

In this paper we review the results of the recent studies 
on the spectra of heavy quarkonia
(bottomonium, charmonium and $B_c$ states) \cite{bsv1,connection}.
We focus mainly on the bottomonium spectrum below; the
results of the charmonium and $B_c$ spectra will be summarized 
at the end of the paper.

For over 20 years, major theoretical approaches to
the charmonium and bottomonium spectroscopy have been those based on
various phenomenological potential models.
In each of the models a non-relativistic Hamiltonian,
\bea
\hat{H} = \frac{\vec{p}\, ^2}{2 m_r} + V_{\rm pheno}(r) ,
\eea
is assumed and a phenomenological potential $ V_{\rm pheno}(r)$
is determined such that the observed
quarkonium spectra (and some other physical observables) are 
reproduced, starting from a simple ansatz for the form of the
potential.
The determined potentials in all models have
more or less similar slopes in the
range $0.5~{\rm GeV}^{-1} \simlt r \simlt 5~{\rm GeV}^{-1}$, which
may be represented by a logarithmic potential
$\propto \log r + {\rm const}$.
These phenomenological-model approaches have successfully elucidated
nature of the quarkonium systems, such as their leptonic widths
and transitions among different levels, besides reproducing
the energy levels.
See e.g.\ Ref.\cite{eq} for a most recent analysis based on the
potential models.
An apparent deficit of these approaches is, however, a
difficulty in relating phenomenological parameters to the fundamental
parameters of QCD.

\begin{figure}[tbp]
  \begin{minipage}{6cm}\centering
    \includegraphics[width=5cm]{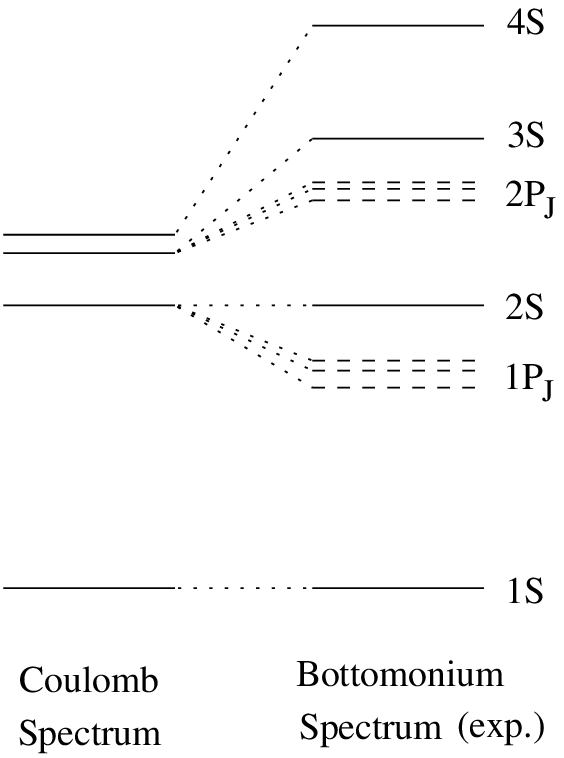}
  \end{minipage}
\hspace{5mm}
  \begin{minipage}{10.0cm}\centering 
    \includegraphics[width=9cm]{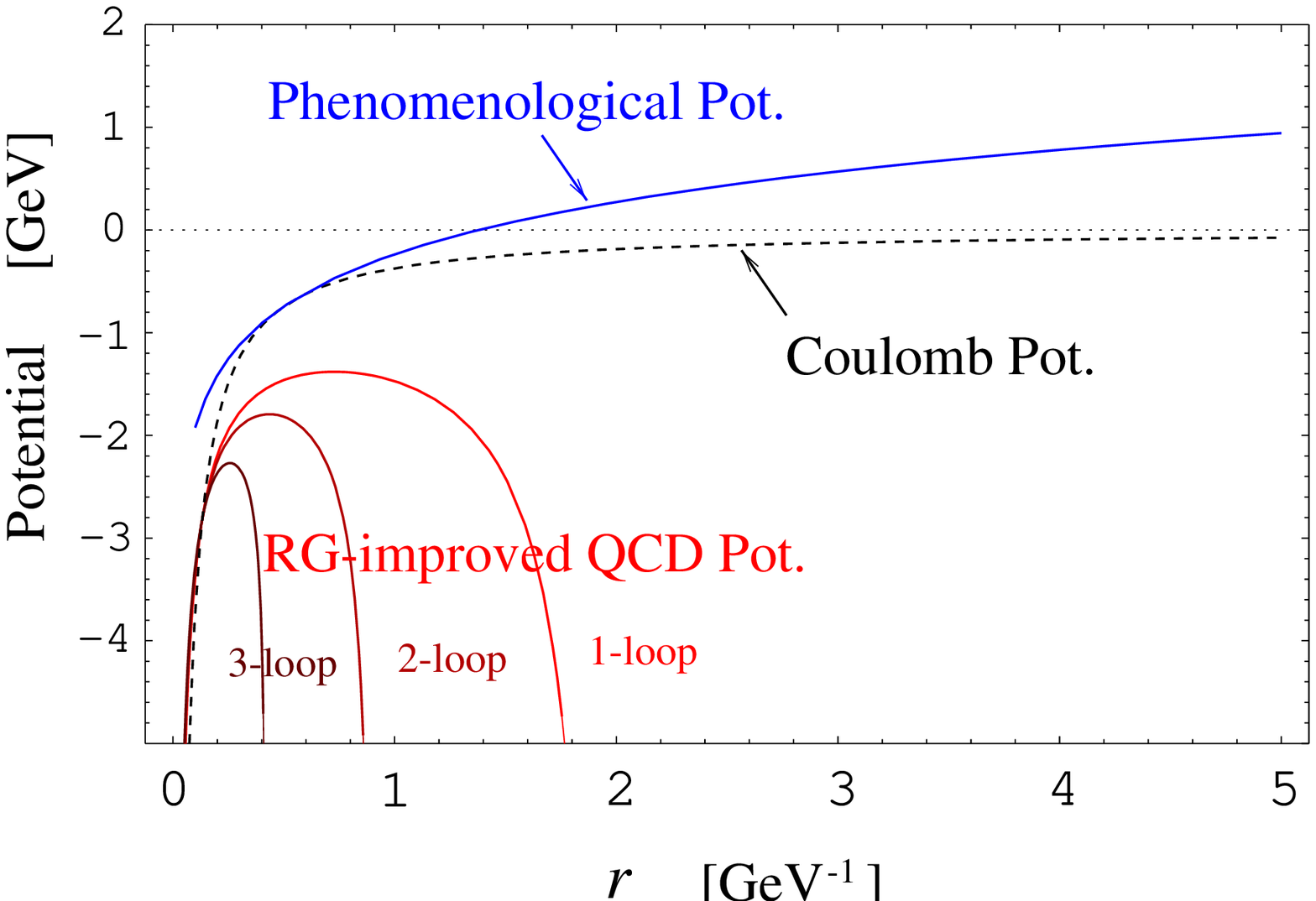}
  \end{minipage}
\vspace{8mm}
  \\
\hspace{2.3cm}
(a)
\hspace{8.6cm}
(b)
\caption{\footnotesize
(a) A comparison of the Coulomb spectrum and the observed bottomonium 
spectrum.
The parameters $m_b$ and $\alpha_S$ of the Coulomb spectrum are
adjusted to reproduce the observed $1S$ and $2S$ levels.
(b) A comparison of the renormalization-group-improved QCD potential,
the Coulomb potential and a typical phenomenological potential.
\label{intro}
}
  \hspace*{\fill}
\end{figure}

The reason why people had to resort to phenomenological models is
because the theory of non-relativistic boundstates,
which has been successful in describing the spectra of the QED boundstates,
failed to reproduce the charmonium and bottomonium spectra in QCD.
Within this theory, the quarkonium states in the leading approximation 
are described by the Hamiltonian of non-relativistic
quantum mechanics with a Coulomb potential
\bea
V_{\rm C}(r) = - \, C_F \frac{\alpha_S}{r} .
\label{Coulombpot}
\eea
It stems from one gluon exchange between the quark and antiquark,
where $C_F = 4/3$ is a color factor.
Compare the observed bottomonium spectrum and the Coulomb spectrum
shown in Fig.~\ref{intro}(a).
In the Coulomb spectrum, the level spacing between consecutive
$nS$ states decreases rapidly as $1/n^2$.
On the other hand in the bottomonium
spectrum the level spacing appears roughly constant;
furthermore, the separations between the $S$ and $P$ states as well
as the fine splittings among $nP_J$ states are sizable.
Thus, the level structures look qualitatively very different.
Including higher order corrections, the Coulomb potential changes to
the QCD potential, which is given roughly by replacing 
$\alpha_S$ in the numerator of Eq.~(\ref{Coulombpot}) with the
running coupling constant evaluated at scale
$1/r$, i.e.\ $\alpha_S(\mu=1/r)$.
Accordingly the QCD potential is bent {downwards} at long distances 
as compared to the Coulomb potential.
As can be seen in Fig.~\ref{intro}(b), the QCD potential becomes
singular at a fairly short distance.
At $r \simgt 0.2~{\rm GeV}^{-1}$, the QCD potential is poorly convergent
as the higher order corrections are included.
There seems to be
no chance that it explains the phenomenologically determined potentials,
which deviate from the Coulomb potential in the 
upward (i.e.\ opposite) direction.
The large discrepancy of the QCD potential and phenomenological
potentials, as well as the poor convergence of the QCD potential,
have been thought as indications of
large non-perturbative effects inherent in the heavy quarkonium systems.
In fact the difference between a typical phenomenological potential
and the Coulomb potential tends to be a linearly rising potential at
distances $r \simgt 1~{\rm GeV}^{-1}$,
suggesting confinement of quarks.

During the last few years, the theory of non-relativistic QCD boundstates
has developed remarkably.
There were two important developments:
(1) The complete next-to-next-to-leading order corrections to the
energy levels have been computed \cite{yn3,py,my,bcbv,hoangcm}.
Also, some of the (non-trivial)
next-to-next-to-next-to-leading order corrections have been calculated
\cite{n3lo,mce,ks}.
(2) The renormalon cancellation in the energy levels was discovered
\cite{renormalon1,renormalon2}.
The upshot is that convergence of the
perturbative expansions of the energy levels improves drastically
if we express the levels in terms of the $\overline{\rm MS}$ mass
instead of the pole mass of a quark.
These theoretical developments enabled accurate perturbative
computations of the quarkonium energy levels.
Let us demonstrate the improvement of convergence
for the $\Upsilon (1S)$ and $\Upsilon(2S)$ states:
\\
\\
\begin{tabular}{lp{15cm}}
$~\bullet~\Upsilon(1S)$: &
For $\mu = 2.49$~GeV, $\alpha_S(\mu)=0.274$,
$m_b^{\overline{\rm MS}}(m_b^{\overline{\rm MS}})
=4.20$~GeV/$m_{b,{\rm pole}}=4.97$~GeV,
\end{tabular}
\bea
M_{\Upsilon(1S)} &=& 
9.94 - 0.17 - 0.20 - 0.30 ~~{\rm GeV}
~~~~~
\mbox{(Pole-mass scheme)}
\\
&=&
8.41 + 0.84 + 0.20 + 0.013~{\rm GeV}
~~~~~
\mbox{($\overline{\rm MS}$-scheme)} .
\eea
\\
\begin{tabular}{lp{15cm}}
$~\bullet~\Upsilon(2S)$: &
For $\mu = 1.09$~GeV, $\alpha_S(\mu)=0.433$,
$m_b^{\overline{\rm MS}}(m_b^{\overline{\rm MS}})
=4.20$~GeV/$m_{b,{\rm pole}}=4.97$~GeV,
\end{tabular}
\bea
M_{\Upsilon(2S)} &=& 
9.94 - 0.10 - 0.19~ - 0.45 ~~{\rm GeV}
~~~~~
\mbox{(Pole-mass scheme)}
\\
&=&
8.41 + 1.46 + 0.093 + 0.009~{\rm GeV}
~~~~~
\mbox{($\overline{\rm MS}$-scheme)} .
\eea
As can be seen, 
when the pole mass is used the series are not converging, whereas
the series show healthy convergent behaviors when the $\overline{\rm MS}$
mass is used.
As important applications of these developments
up to date, the theory enabled precise 
determinations of the $\overline{\rm MS}$-mass of the bottom 
quark \cite{py,pp,my,upsilonmass}
and (in the future) of the top quark \cite{topcollab} from (mainly) 
the energy levels of the lowest-lying states. 
The main uncertainty comes, in the bottomonium case, 
from the (essentially) unknown non-perturbative contributions. These are 
generally claimed 
to be around 100 MeV and ultimately set the precision of the prediction. 

Based on the above theoretical developments we analyze the consistency 
between the whole level structure of the quarkonium
as predicted by the boundstate
theory (perturbative QCD) and that of the experimental data.
Now that we can make accurate predictions, we can
extract upper bounds on the non-perturbative contributions to the
energy levels by comparing the perturbative predictions, 
at the current best accuracy, with the experimental data.
It turns out that non-perturbative effects should be much smaller than
what have been believed conventionally.

The paper is organized as follows.
In Sec.~2 we review briefly the necessary theoretical framework.
We analyze the bottomonium spectrum numerically in Sec.~3 and
estimate the errors of the theoretical predictions in Sec.~4.
We interpret the result and provide a new physical picture of
the bottomonium states in Sec.~5.
Then we discuss a connection between our approach and the 
phenomenological potential-model approaches in Sec.~6.
We draw conclusions in Sec.~7.

\section{Theoretical Framework}

\subsection{$1/c$ expansion}

We state briefly the theoretical framework used in contemporary
calculations of the spectrum of heavy quarkonia.
The problem is reduced to a quantum mechanical one as follows.
We first compute the quark-antiquark (off-shell) 
scattering amplitude in ordinary perturbative QCD.
We then determine a quantum mechanical Hamiltonian such that the
quark-antiquark scattering amplitude computed within
quantum mechanics matches the former amplitude order by
order in expansion in $1/c$ (inverse of the speed of light):
\bea
\hat{H} =  \hat{H}_0 + \frac{1}{c} \, \hat{H}_1
+ \frac{1}{c^2} \, \hat{H}_2 + \cdots .
\label{nrhamiltonian}
\eea
The expansion is a double expansion in
$\alpha_S = g_S^2/(4\pi \hbar c)$ and $\beta = v/c$.
Then we solve the non-relativistic Schr\"odinger equation
\bea
\hat{H} \, \psi_n (r) = E_n \, \psi_n (r)
\label{schroedingereq}
\eea
to determine the boundstate wave functions and energy spectrum,
order by order in $1/c$ expansion.
Provided the quark and antiquark inside the heavy quarkonium are
non-relativistic, the expansion in $1/c$ leads to a
reasonable systematic approximation.

It should be noted that this procedure parallels a more
familiar procedure for the calculation of the masses of one-particle states.
The matrix element of the above Hamiltonian
$\bra{\vec{p}} \hat{H} \ket{\vec{p}\, '}$ is,
in the language of perturbative QCD, the sum of two-particle
irreducible diagrams.\footnote{
This is true with respect to the diagrams in the time-ordered
(old-fashioned) perturbation theory \cite{hasebe}.
}
One may then compare Eq.~(\ref{schroedingereq}) 
with e.g.\ the computation of the photon and $Z$ boson
masses by solving the eigenvalue equation
\bea
s \vc{1} - 
\left(
\begin{array}{cc}
\Pi_{\gamma\gamma}(s) & \Pi_{\gamma Z}(s) \\
\Pi_{Z \gamma}(s) & \Pi_{ZZ}(s) 
\end{array}
\right) = 0 ,
\nonumber
\eea
where $\Pi_{ij}(s)$ represent the sums of one-particle irreducible diagrams
[$\Pi_{ZZ}(s) $ includes $M_{Z}^{\rm (tree)}$].

Presently the Hamiltonian is known up to ${\cal O}(1/c^2)$ in 
the Coulomb gauge
\cite{py,htmy}:
\bea
\rule[-6mm]{0mm}{6mm}
\hat{H}_0 &=& \frac{\vec{p}\,^2}{m} \,
{- \, C_F \,\frac{\alpha_S}{r}} , 
\label{h0}
\\ \rule[-8mm]{0mm}{6mm}
\hat{H}_1 &=& 
{- \, C_F \,\frac{\alpha_S}{r} \cdot
\biggl( \frac{\alpha_S}{4\pi} \biggr) \cdot
\biggl\{ \beta_0 \, \log ( \mu'^2 r^2 ) } + a_1 \biggr\}
, 
\\
\hat{H}_2 &=& - \frac{\vec{p}\,^4}{4m^3} \,
{- \, C_F \,\frac{\alpha_S}{r}\cdot
\biggl( \frac{\alpha_S}{4\pi} \biggr)^2 \cdot
\biggl\{ \beta_0^2 \,[ \log^2 ( \mu'^2 r^2 ) + \frac{\pi^2}{3}] 
}
+ (\beta_1+2\beta_0 a_1)\log ( \mu'^2 r^2 ) + a_2
\biggr\}
\nonumber \\ &&
+ \frac{\pi C_F \alpha_S}{m^2} \, \delta^3(\vec{r})
+\frac{3 C_F \alpha_S}{2m^2r^3} \, \vec{L}\cdot\vec{S}
- \frac{C_F \alpha_S}{2m^2r} \biggl(
\vec{p}\,^2 + \frac{1}{r^2} r_i r_j p_j p_i \biggr)
- \frac{C_A C_F \alpha_S^2}{2mr^2}
\nonumber \\ &&
- \frac{C_F \alpha_S}{2m^2}
\biggl\{ \frac{S^2}{r^3} - 3 \frac{(\vec{S}\cdot\vec{r})^2}{r^5}
- \frac{4\pi}{3}(2S^2-3) \delta^3(\vec{r}) 
\biggr\} , \label{h2}
\eea
where $m$ denotes the pole mass of the quark;
$\alpha_S \equiv \alpha_S(\mu)$;
$C_F = 4/3$, $C_A=3$ are color factors; 
$\mu' = \mu \, e^{\gamma_E}$.
The lowest-order Hamiltonian $\hat{H}_0$
is nothing but that of 
two equal-mass particles interacting via the
Coulomb potential.

\subsection{Renormalon cancellation}

In addition to the first three terms $\hat{H}_0$, $\hat{H}_1$, $\hat{H}_2$ 
of the Hamiltonian, part of the higher order terms
$\hat{H}_n$ are known.
Specifically, these are part of the static QCD potential $V_{\rm QCD}(r)$.
From their analysis,
it has been known \cite{al} that the series expansion of
$V_{\rm QCD}(r)$ in $\alpha_S(\mu)$
diverges rapidly at high orders, and that this results in
an uncertainty of $V_{\rm QCD}(r)$ of order $\Lambda_{\rm QCD}$ 
even within perturbative QCD.
This problem is referred to as the renormalon problem.
\begin{figure}[tbp]
  \hspace*{\fill}
  \begin{minipage}{5.0cm}\centering
    \hspace*{-2.3cm}
    \includegraphics[width=8cm]{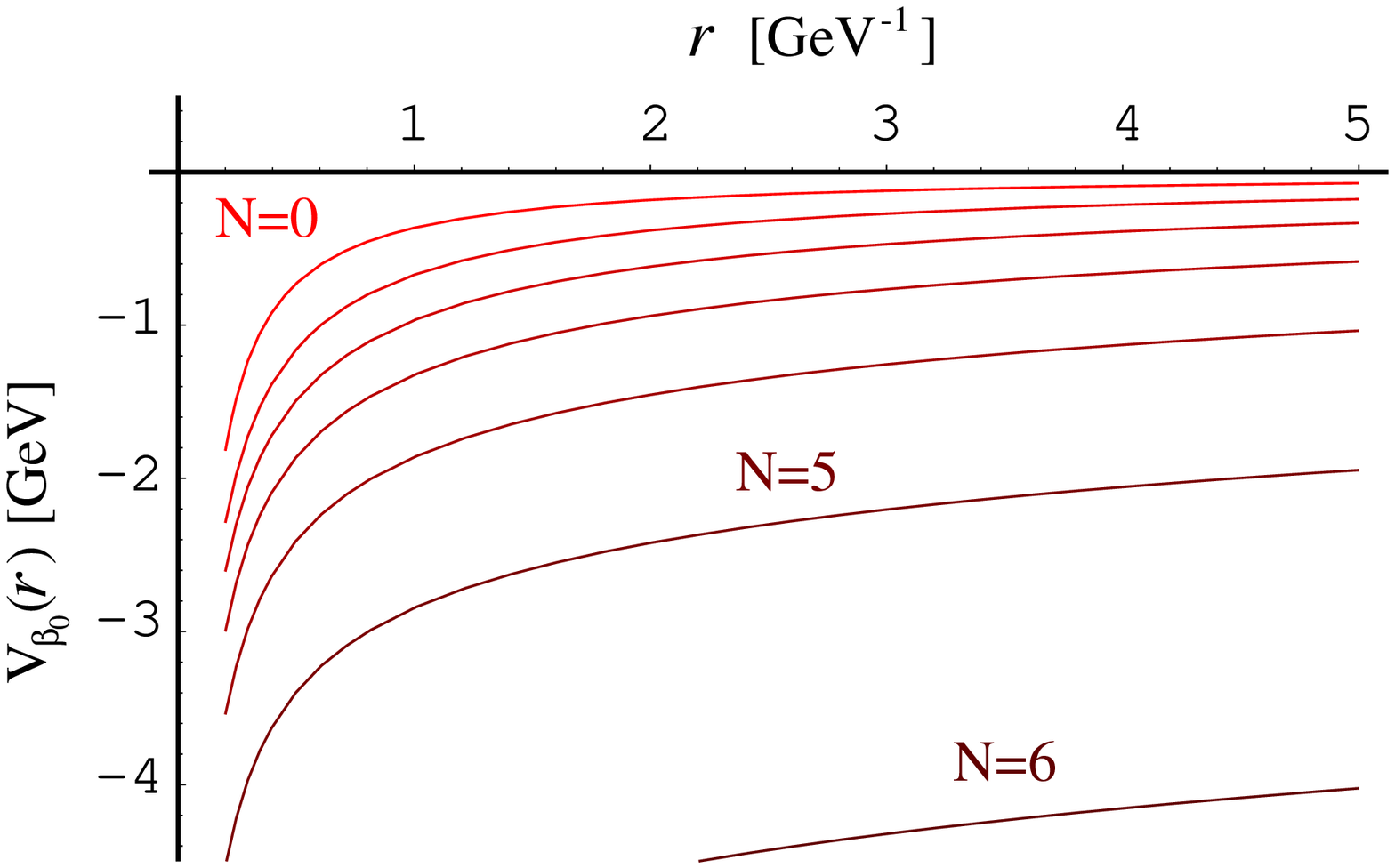}\vspace{5mm}
\hspace*{-1cm}(a)
  \end{minipage}
  \hspace*{\fill}
  \begin{minipage}{5.0cm}\centering
    \hspace*{-1cm}
    \includegraphics[width=8cm]{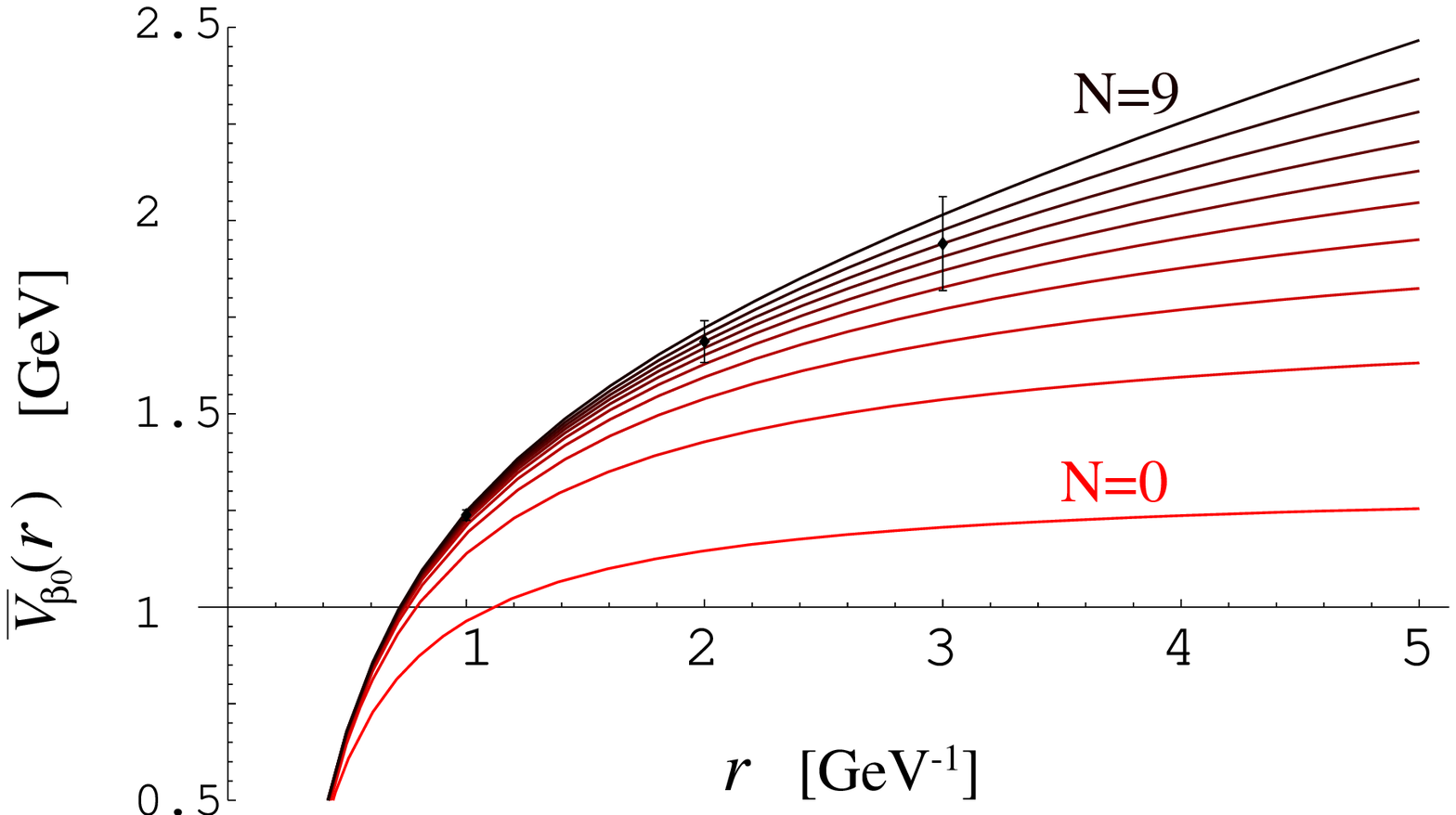}\vspace{9mm}
    \hspace*{1.2cm}(b)
  \end{minipage}
  \hspace*{\fill}
  \\
  \hspace*{\fill}
\caption{\footnotesize
The QCD potential in the large-$\beta_0$ approximation
truncated at $O(\alpha_S^{N+1})$ term.
We set $\mu=2.49$~GeV, $n_l=4$ and $\alpha_S(\mu)=0.273$
[corresponding to $\alpha_S^{(5)}(M_Z)=0.1181$].
(a) Before subtraction of the leading renormalon.
(b) After subtraction of the leading renormalon.
These figures are taken from \cite{connection}.
      \label{large-beta0}
}
  \hspace*{\fill}
\end{figure}
This is demonstrated in Fig.~\ref{large-beta0}(a), where 
the QCD potential in the so-called ``large-$\beta_0$ approximation"
is shown  
up to the $(N \! + \! 1)$-th term, $\sum_{n=0}^N V^{(n)}_{\beta_0}(r)$,
for $N=0,1,2,\dots$ and $n_l=4$.
We see that the higher order corrections are indeed large and
almost constant (independent of $r$).

It was found \cite{renormalon1,renormalon2} 
that the leading renormalons contained in the pole mass \cite{bbb}
and in the QCD potential \cite{al} cancel in the total energy of a static 
quark-antiquark pair
if the pole mass $m_{\rm pole}$ is expressed in terms of
the $\overline{\rm MS}$ mass:
\begin{eqnarray}
&&
E_{\rm tot}(r) \equiv 2 m_{\rm pole} + V_{\rm QCD}(r) ,
\label{total-ene}
\\
&&
V_{\rm QCD}(r) \simeq - \int 
\frac{d^3\vec{q}}{(2\pi)^3} 
\, e^{i \vec{q} \cdot \vec{r}} \, 
C_F \frac{4\pi\alpha_S(q)}{q^2} ,
\label{renormalon-pot}
\\
&&
m_{\rm pole} \simeq m_{\overline {\rm MS}}(\mu ) +
\frac{1}{2} {\hbox to 18pt{
\hbox to -5pt{$\displaystyle \int$} 
\raise-15pt\hbox{$\scriptstyle q< \mu$} 
}}
\frac{d^3\vec{q}}{(2\pi)^3} \, 
C_F \frac{4\pi\alpha_S(q)}{q^2} ,
\label{renormalon-mass}
\end{eqnarray}
where $q = |\vec{q}|$.
The above equations show that
the potential $V_{\rm QCD}(r)$ is essentially the Fourier transform of the
Coulomb gluon propagator exchanged between quark and antiquark, and that
the difference of $m_{\rm pole}$ and $m_{\overline {\rm MS}}$ is 
essentially the infrared portion of the quark self-energy.
The renormalon contributions originate from the infrared region,
$q \sim \Lambda_{\rm QCD}$, of the loop integrations, where
the running coupling constant $\alpha_S(q)$ becomes large.
Namely the infrared gluons cause bad behaviors of the perturbation
series at high orders.
The signs of the renormalon contributions are opposite
between $V_{\rm QCD}(r)$ and $m_{\rm pole}$ because the color charges are
opposite between quark and antiquark while the self-enregy is
proportional to the square of a same charge.
Their magnitudes differ by a factor of two because
both the quark and antiquark propagator poles contribute in the
calculation of the potential whereas only one of the two contributes
in the calculation of the self-energy.
Expanding the Fourier factor $e^{i \vec{q} \cdot \vec{r}}$ 
in a Taylor series for small $\vec{q}$, the leading
renormalon contributions cancel in $E_{\rm tot}(r)$.

As a result of this cancellation, 
the series expansion of the total energy in $\alpha_S(\mu)$ converges
better if we use the $\overline{\rm MS}$ mass instead of
the pole mass.
In order to demonstrate the improvement of convergence, we show
in Fig.~\ref{large-beta0}(b) 
the QCD potential in the large-$\beta_0$ approximation
[up to the $(N \! + \! 1)$-th term] after 
the leading renormalon is subtracted at each order of $\alpha_S(\mu)$.
One sees that the series expansion of the potential has become 
much more convergent as compared to Fig.~\ref{large-beta0}(a).
Note that the higher-order corrections raise the potential
at long distances after subtraction of the renormalon contribution.
Some important aspects are:
\begin{itemize}
\item
The pole mass of a quark is ill-defined beyond perturbation theory.
It can be determined only when the quark can propagate 
an infinite distance.
Generally accepted belief is that when quark and antiquark are separated
beyond a distance $\sim \Lambda_{\rm QCD}^{-1}$ the color flux 
is spanned between the two charges due to non-perturbative effects 
and the free quark picture is no longer valid.
On the other hand, the total
energy (or the mass) of a quarkonium, which is
a color-singlet state, is  physically meaningful.
A color-singlet state can propagate for a long time and the notion of its
mass is not limited by the hadronization scale.
\item
When the size of a color-singlet system is much smaller than
$\Lambda_{\rm QCD}^{-1}$, infrared gluons with wavelengths 
$\Lambda_{\rm QCD}^{-1}$ cannot couple
to color sources inside the system ---
such a picture is naturally described by the bare QCD Lagrangian.
Hence, if we use $m_{\overline {\rm MS}}$,
which is more closely related to the bare mass than $m_{\rm pole}$ is,
contributions from the infrared gluons vanish in $E_{\rm tot}(r)$.
\end{itemize}
See e.g.\ \cite{sumino} for an introductory review of the renormalons
in the heavy quarkonium states.

\section{Bottomonium Spectrum: A Numerical Analysis}

In this section we examine the bottomonium spectra numerically
\cite{bsv1}.
According to the formalism explained in the previous section,
the energy levels are computed analytically as functions of 
$\alpha_S(\mu)$, $\mu$ and
$\overline{m}_b \equiv m_b^{\overline{\rm MS}}(m_b^{\overline{\rm MS}})$ 
(the $b$-quark
$\overline{\rm MS}$ mass renormalized at the $\overline{\rm MS}$-mass scale).
The dependence on the scale $\mu$ arises from truncation of the
series expansion at finite order.
In this section we set the number of massless flavors as $n_l=4$ and
neglect the effects of the non-zero charm quark 
mass in the $b\bar{b}$ systems. 

The algorithm of our calculations goes as follows.
\begin{enumerate}
\item
We take the strong coupling constant with the present world average
value as input,\footnote{
We evolve the coupling and match 
it to the coupling of the theory with $n_l=4$ 
via 4-loop running. 
}
$\alpha_S^{(5)}(M_Z) = 0.1181 \pm 0.0020$ \cite{pdg}.
\item
The scale $\mu$ is fixed for each state $X$ from the minimal sensitivity
condition:
\bea
\left.\frac{d}{d\mu} E_{X}(\mu, \alpha_S(\mu),\overline{m}_b)
\right|_{\mu = \mu_X} = 0 .
\label{scalefix}
\eea
\item
We fix $\overline{m}_b$ from the mass of the vector ground state:
\bea
E_{\Upsilon(1S)} (\mu_{1S}, \alpha_S(\mu_{1S}),\overline{m}_b) 
&=& E_{\Upsilon(1S)}^{exp} = 9.460 \, \hbox{GeV} .
\label{mbfix}
\eea
\end{enumerate}

A comment regarding the scale determined by the
scale-fixing prescription Eq.~(\ref{scalefix}) 
is in order.
We find that, for most of the bottomonium states,
the convergence properties of the series become optimal at
$\mu \simeq \mu_X$, and
that the scale becomes close to the inverse of the physical size of the 
boundstate $X$. 
If the scale fixed by Eq.~(\ref{scalefix}) evidently does not fulfill 
these conditions, then the theoretical predictions obtained in this way 
will be considered {\it unreliable}.
We will show that this typically happens for the higher levels,
where the coupling constant becomes bigger than one.

\begin{figure}[p] 
\begin{center}
    \includegraphics[width=10cm]{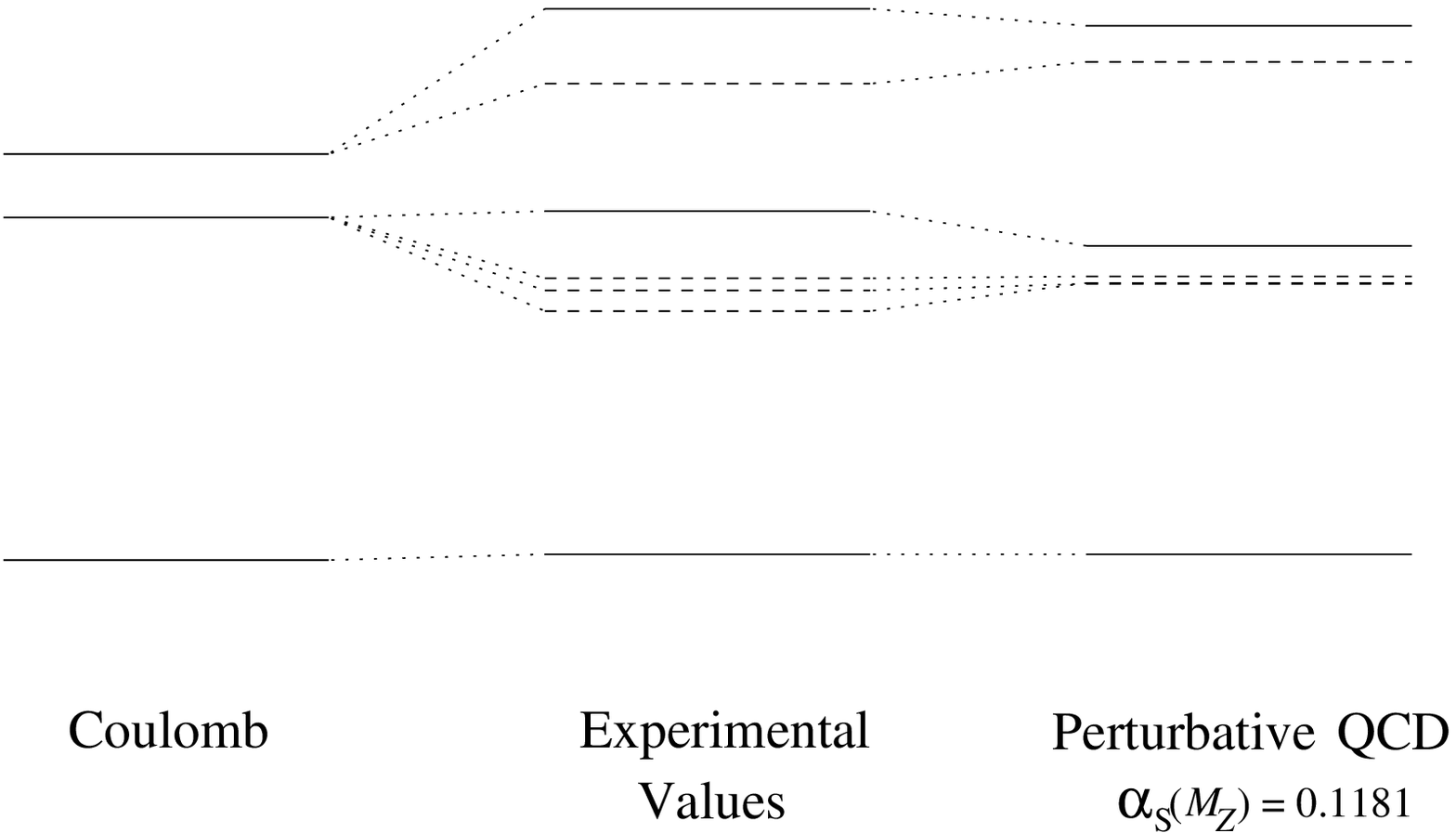}
\vspace{7mm}
\\
\hspace{-20mm}
    \includegraphics[width=18cm]{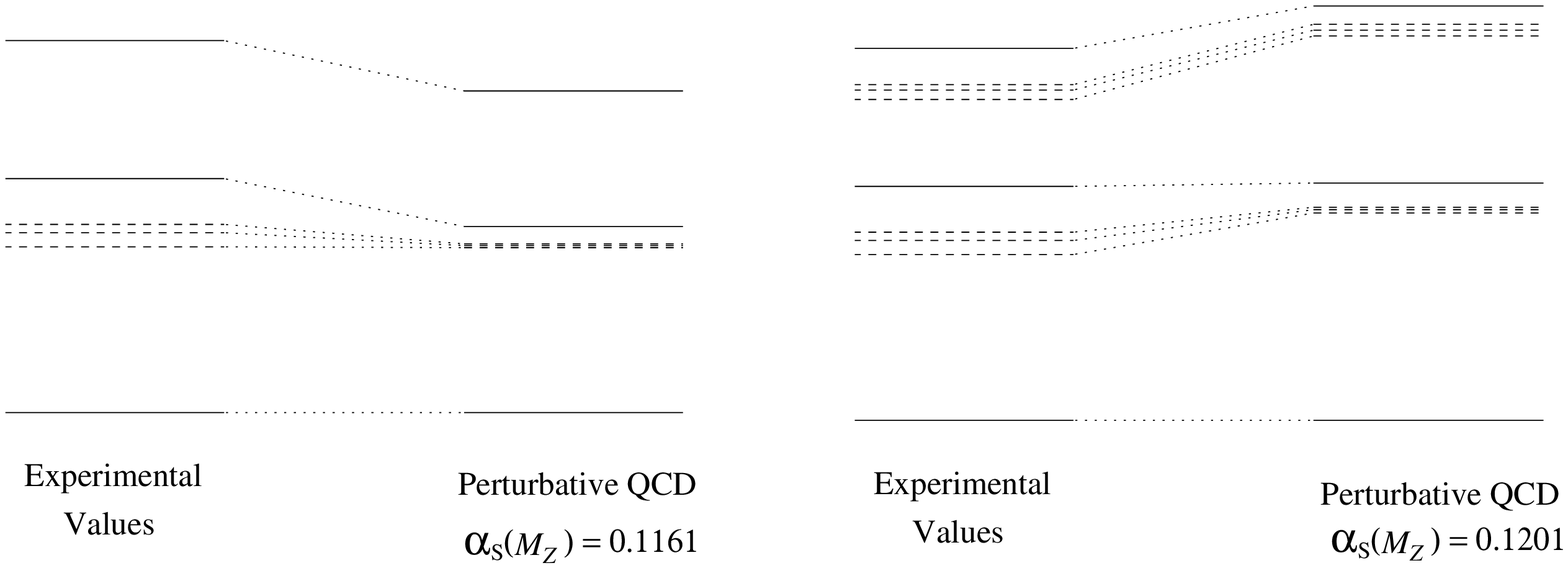}
\end{center}
\caption{\footnotesize
Comparisons of the theoretical predictions and the experimental data for
the bottomonium spectrum.
The input values of the theoretical predictions
are $\alpha_S^{(5)}(M_Z)=0.1181$,
0.1161 and 0.1201.
We set $m_c=0$ and the number of massless flavors as $n_l=4$.
The solid and dashed lines represent the $S$- and $P$-states,
respectively.
We show only those levels which we can compute reliably.
\label{bottomoniumspect}
}
\end{figure}

The $b$-quark $\overline{\rm MS}$ mass fixed by Eq.~(\ref{mbfix}) is
given by
\begin{table}[h]
\begin{center}
\begin{tabular}{c||c|c|c}
\hline
$\alpha_S^{(5)}(M_Z)$ & 0.1161 & 0.1181 & 0.1201 \\
\hline
$\overline{m}_b$  & 4.221~GeV  & 4.203~GeV & 4.184~GeV \\
\hline
\end{tabular}
\label{tab:mbfix}
\end{center}
\end{table}
\\
Using these masses as inputs and Eq.~(\ref{scalefix}), we can calculate the 
energy levels of other observed quarkonium states. These are shown in
Figs.~\ref{bottomoniumspect}.
Only those levels which can be predicted reliably are displayed.
The level spacings become wider for larger $\alpha_S^{(5)}(M_Z)$,
which is consistent with our naive expectation. 
If we take an average of the $S$-wave and $P$-wave 
levels corresponding to each principal quantum number $n$, the 
theoretical predictions 
with $\alpha_S^{(5)}(M_Z)=0.1181$ reproduce the experimental values fairly 
well.
On the other hand, 
the predictions for the $S$--$P$ splittings and the fine splittings 
are smaller than the experimental values.

\begin{table}[t]
\begin{center}
\begin{tabular}{c|r|r|c|l|r|l|l|l}
\hline
State $X$ &$E_X^{exp}~~$ &$E_X~~~$ & $E_X^{exp}-E_X$ &
$E_X^{(1)}$ &$E_X^{(2)}$ &$E_X^{(3)}$ &~~$\mu_X$ 
&$\alpha_S(\mu_X)$\\ 
\hline
$\Upsilon(1^3S_1)$  &  9.460~~ & 9.460~ & 0        & 0.837 & 0.204  & 0.013 & 2.49 & 0.274  \\
$\Upsilon(1 ^3P_0)$ &  9.860~~ & 9.905~ & $-0.045$ & 1.38  & 0.115  & 0.003 & 1.18 & 0.409  \\
$\Upsilon(1 ^3P_1)$ &  9.893~~ & 9.904~ & $-0.011$ & 1.40  & 0.098  & 0.002 & 1.15 & 0.416  \\
$\Upsilon(1 ^3P_2)$ &  9.913~~ & 9.916~ & $-0.003$ & 1.42  & 0.086  & 0.003 & 1.13 & 0.422  \\
$\Upsilon(2 ^3S_1)$ & 10.023~~ & 9.966~ & $+0.057$ & 1.46  & 0.093  & 0.009 & 1.09 & 0.433  \\
$\Upsilon(2 ^3P_0)$ & 10.232~~ &10.268 ~& $-0.036$ & 2.37  & $-0.66~\,$ & 0.15&0.693 & 0.691  \\
$\Upsilon(2 ^3P_1)$ & 10.255~~ &10.316$^\sharp$ & ~$-0.061$$^\sharp$ & 3.97 & $-3.56~\,$ 
& 1.50 &0.552$^\sharp$ & 1.20  \\
$\Upsilon(2 ^3P_2)$ & 10.268~~ &10.457$^\sharp$ & ~$-0.189$$^\sharp$ & 4.55 & $-5.03~\,$ 
& 2.53 & 0.537$^\sharp$ & 1.39  \\
$\Upsilon(3 ^3S_1)$ & 10.355~~ &10.327~ & $+0.028$ & 2.34 & $-0.583$& 0.163&0.698 & 0.684  \\
$\Upsilon(4 ^3S_1)$ & 10.580~~ &11.760$^\sharp$ & ~$-1.180$$^\sharp$ & 5.45 & $-6.47~\,$ & 4.38 
&0.527$^\sharp$ & 1.61\\
\hline
\end{tabular}
\end{center}
\vspace{-3mm}
\caption{\footnotesize
Comparisons of the theoretical predictions of perturbative QCD and the experimental data for $\alpha_S(M_Z)=0.1181$.
All dimensionful numbers are in GeV unit.
}
\label{table:spectra}
\end{table}
\begin{table}[]
\begin{center}
\begin{tabular}{c|l||r|r|r|r|r|r|r}
\hline
\multicolumn{2}{c||}{} & 
\multicolumn{6}{c}{~~Estimates of higher-order corrections}\\
\cline{3-9}
\multicolumn{2}{c||}{} 
& (i)~~ & (ii)~~ & (iii)\, & (iv) \, & (v) & $\pm$max & $\mu = 2 \mu_X$\\
\hline
\multicolumn{2}{c||}{$\delta \overline{m}_b$} 
&$-2$~ & +1~ & 0~ & $\pm 7$~ & +5 &$\pm 7$~~ & +16~~\\
\hline
& $\Upsilon(1 ^3P_0)$ &  +15~ & $-14$~ & 0~ & $\pm 3$~ & & $\pm 15$~ & $-53$~~\\
& $\Upsilon(1 ^3P_1)$ & +22~ & $-8$~ & +7~ & $\pm 2$~ & & $\pm 22$~ & $-48$~~\\
& $\Upsilon(1 ^3P_2)$ & +16~ & $-16$~ & 0~ & $\pm 3$~ & & $\pm 16$~ & $-54$~~\\
& $\Upsilon(2 ^3S_1)$~ & +18~ & $-19$~ & $-1$~ & $\pm 9$~ & & $\pm 19$~ & $-73$~ \\
$\delta E_X$ & $\Upsilon(2 ^3P_0)$ &+33~ & $-57$~ & $-21$~ & $\pm 150$~ & & $\pm 150$~ & $-120$~~\\
& $\Upsilon(2 ^3P_1)$ & $-4$~ 
& $+1637^\sharp$ & $-66^\sharp$ & $\pm 1500^\sharp$~& & $\pm 1637^\sharp$ 
& $-97$~~\\
& $\Upsilon(2 ^3P_2)$ & $-136$~ 
& $+2136^\sharp$ &$-206^\sharp$ & $\pm 2530^\sharp$~& &  $\pm 2530^\sharp$ 
& $-229$~~\\
& $\Upsilon(3 ^3S_1)$ & +37~ & $-63$~ &$-36$~ 
& $\pm 163$~ & & $\pm 163$~ & $-161$~~\\
& $\Upsilon(4 ^3S_1)$ & $-639^\sharp$ 
& $+2936^\sharp$ &$-1425^\sharp$ & $\pm 4380^\sharp$& & $\pm 4380^\sharp$ 
& $-1361$~~\\
\hline
\end{tabular}
\end{center}
\caption{\footnotesize
Variations of the theoretical predictions of Tab.~\ref{table:spectra} 
when the uncertainties 2) discussed in Sec.~\ref{s4} are separately
taken into account. 
All dimensionful numbers are in MeV unit.
Those values corresponding to the unreliable theoretical predictions 
are marked with sharps. 
The input parameter is taken as $\alpha_S(M_Z)=0.1181$.
The column ``$\pm$max'' lists $\pm$max$\{|$(i)$|$, $|$(ii)$|$, 
$|$(iii)$|$, $|$(iv)$|$, $|$(v)$|\}$.
In the last column we write conservative estimates with the scale
choice $\mu = 2\mu_X$.}
\label{error}
\end{table}
Also the numerical values of the 
series expansions of the energy levels are shown 
in Tab.~\ref{table:spectra} in the case $\alpha_S^{(5)}(M_Z)=0.1181$. 
In this case,
the series expansions for the $1^3S_1$, $1^3P_J$, $2^3S_1$, 
$2^3P_0$ and $3^3S_1$ bottomonium states converge well. 
For these states the differences of the theoretical predictions and the 
experimental data are typically  $30$--$60$ MeV.
Convergence of the series expansions is poor for the $2P_1$, $2P_2$ and $4S$ 
states. 
We consider that the theoretical predictions for these levels are unreliable 
and marked the corresponding theoretical predictions 
with sharps ($\sharp$).
The differences $E_X^{exp}-E_X$ are rather large 
for the states with sharps. 
Notice that for these states the corresponding $\alpha_S(\mu_X)$ 
becomes larger than one, indicating a breakdown of the perturbative series.
Generally, for states, which we consider reliably calculable, the scale 
dependence decreases as we include more terms of the perturbative series.
For states, whose predictions we consider unreliable, the series becomes 
much more convergent 
if we would choose a scale different from (typically larger than) $\mu_X$.

\section{Error Estimates} 
\label{s4}
Besides non-perturbative corrections, there are two kinds of
uncertainties in our theoretical predictions for the bottomonium spectra.
These are listed below and in Tab.~\ref{error}. 
Also these examinations indicate that the theoretical predictions
for the $2^3P_1$, $2^3P_2$ and $4^3S_1$ bottomonium states are quite unstable, 
while the prediction for the $2^3P_0$ state is at the boundary. 

\begin{itemize}
\item[1)]{{\it Charm Mass Effects.}
We have also made an analysis of the bottomonium spectrum including 
finite charm mass effects.  
We will report the full results in a separate paper \cite{bsv2}. 
Here we only summarize some of the qualitative features of the 
effects and include them as a part of the uncertainties of our present 
analysis.
The corrections to the reliable predictions turn out to be positive and 
to become larger for higher states, ranging up to $\sim 200$ MeV.
Much of the effects, however,  can be cancelled by decreasing the input
$\alpha_S^{(5)}(M_Z)$ within its present uncertainty.\footnote{
We note that the sensitivities of the higher levels
to a variation of $\alpha_S^{(5)}(M_Z)$ increase by the charm mass
effects due to the decoupling of the charm quark.}
}
\item[2)]{{\it Uncertainties from Higher-Order Corrections.}
We take the maximum value of the following five estimates as an 
estimate of uncertainties 
from unknown higher-order corrections for each series expansion:
(i) 
The difference between the theoretical predictions computed
using $\alpha_S(\mu)$ as obtained by solving the renormalization-group 
equation 
perturbatively at 4 loops (Eqs.~(3) and (11) of Ref.~\cite{running})
and numerically at 4 loops (the data of Tab.~\ref{table:spectra}).
(Note that the number of energy levels that can be determined in a reliable 
way is larger with the former definition of the running coupling constant. 
Also in that case, reliable predictions turn out to be close to the 
experimental values.)
(ii) 
The difference between the theoretical predictions computed using the 3-loop 
and the 4-loop (as in Tab.~\ref{table:spectra}) running coupling constants,  
fixing $\alpha_S^{(5)}(M_Z) = 0.1181$.
(iii)
The difference between the theoretical estimates obtained by fixing $\mu_X$ 
on the minimum of  $|E^{(3)}|$ 
and the results of Tab.~\ref{table:spectra}, 
obtained by fixing $\mu_X$ via the condition (\ref{scalefix}). 
(iv)
The contribution $\pm|E^{(3)}_X|$ from Tab.~\ref{table:spectra}. 
(v)
For the $1S$ states we consider the ${\cal O}(\alpha_S^5 m)$ corrections 
calculated in the large-$\beta_0$ approximation in \cite{ks}.\\
For comparison, we list more conservative error estimates.
These are the variations of $\overline{m}_{b}$ and $E_X$
when we fix the scale as twice\footnote{
If we fix the scale as half of the minimal sensitivity scale,
$\mu = \mu_X/2$,
the predictions for the $n=2$ bottomonium
states appear to be meaningless, since the
scales are quite close to the infrared
singularity of the runnning coupling constant, and
the predictions for the $n \geq 3$ states do not exist, since the scales
lie below the infrared singularity.
}
of the minimal sensitivity scale:
$\mu = 2 \mu_X$, where $\mu_X$ is determined from Eq.~(\ref{scalefix}).
}
\end{itemize}

\section{Interpretation}
\label{s5}
\clfn
The most non-trivial feature of the present theoretical predictions for the 
bottomonium spectrum is that the level spacings between consecutive $n$'s 
are almost constant,
whereas in the Coulomb spectrum the level spacings decrease as 
$1/n^2$.
Conventionally, this same difference between the Coulomb spectrum
and the observed quarkonium spectra motivated people to construct
various potential models. It is, therefore, imperative to elucidate how
the above perturbative QCD calculation is able to reproduce such a level 
structure.
We will focus on two points:  
(1) the leading renormalon cancellation, which implies that 
infrared physics decouples; 
this is essential to obtain convergent series expansions; 
(2) the meaning of the scale $\mu_X$ 
chosen by the scale-fixing prescription (\ref{scalefix}).

According to the discussion in Sec.~2.2, physical meaning of the
renormalon cancellation can be understood as the decoupling of 
infrared gluons in the computation of the quarkonium energy levels:
The gluons, whose wavelengths are much larger than the size 
of the color-singlet boundstate, cannot couple to it.
In Eqs.~(\ref{renormalon-pot}) and (\ref{renormalon-mass}), the
integrands essentially cancel each other in the region $q \simlt 1/r$.
We may replace $r$ by the size $a_X$ of a boundstate $X$ in 
Eqs.~(\ref{total-ene})--(\ref{renormalon-mass}) and write
\bea
E_X \approx 2 \overline{m} + 
{\hbox to 18pt{
\hbox to -9pt{$\displaystyle \int$} 
\raise-21pt\hbox{$\scriptstyle 1/a_X$} 
\raise-20pt\hbox to -3pt{$\scriptstyle <$}
\raise-24pt\hbox{$\scriptstyle \sim$} 
\raise-21pt\hbox{$\scriptstyle q <\overline{m}$} 
}}
\frac{d^3\vec{q}}{(2\pi)^3} \,
C_F \frac{4\pi\alpha_S(q)}{q^2} .
\label{app-Etot}
\eea
This expression may be made somewhat more accurate.
Let us approximate the energy level as
\bea
&&
E_X \equiv 2 m_{\rm pole} + E_{{\rm bin},X} ,
\\
&&
2m_{\rm pole} \simeq 2 \overline{m} +
{\hbox to 18pt{ \hbox to -5pt{$\displaystyle \int$} 
\raise-19pt\hbox{$\scriptstyle {q}< \overline{m}$} }}
\frac{d^3\vec{q}}{(2\pi)^3} \, C_F \frac{4\pi\alpha_S(q)}{q^2} 
= 2 \overline{m} + \frac{2 C_F}{\pi} \int_0^{\overline{m}} dq \, \alpha_S (q) ,
\label{poleapprox} \\
&& E_{{\rm bin},X} \simeq \bra{X} \frac{\vec{p}^{\, 2}}{m_{\rm pole}}
+V_{\rm QCD} \ket{X}.
\label{binapprox}
\eea 
Here, $V_{\rm QCD}(q) \simeq - C_F 4\pi \alpha_S(q)/q^2$ is 
the QCD static potential in momentum space;
$\ket{X}$ denotes the Coulomb wave function (the zeroth-order 
approximation).
From Eqs.~(\ref{poleapprox}) and (\ref{binapprox}) we obtain 
\bea
E_X &\simeq& 2 \overline{m} + \frac{2C_F}{\pi} \int_0^\infty dq \, 
\alpha_S (q) \, f_X(q) 
+ \bra{X} \frac{\vec{p}^{\,2}}{m_{\rm pole}} \ket{X} 
\\ 
&\simeq& 2 \overline{m} + \frac{2C_F}{\pi} \int_0^\infty dq \, \alpha_S (q) 
\, f_X(q).  
\label{eth3}
\eea
The last approximation follows from the fact that the kinetic energy 
contribution to the bottomonium 
levels turns out to be numerically small\footnote{It is about 
17\% of $E_X^{(1)}$ for the $X=1S$ state, 6\% of $E_X^{(1)}$ for the
$X=2S$ state, 4\% of $E_X^{(1)}$ for the $X=3S$ state. 
Moreover, these contributions tend to cancel each other in the 
level spacings.} 
(notice that this does not contradict the virial theorem, 
since the static potential we are considering here incorporates the running of $\alpha_S$ and therefore 
is not simply of the form $1/r$). 
\begin{figure}[t]
\begin{center}
\vspace{0mm}
\begin{minipage}{8.0cm}\centering
    \includegraphics[width=8cm]{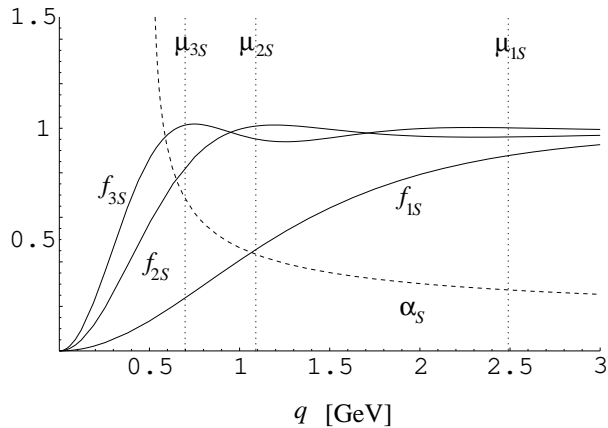}
\vspace{-3mm} \\
  \end{minipage}
\vspace{-7mm}
\end{center}
\caption{\footnotesize
The support functions $f_X(q)$ vs.\ $q$ for $X = 1S$, $2S$ and $3S$ 
(solid lines).
$f_X(q)$ is calculated using $m_{\rm pole}=5$~GeV and a 
different $\alpha_S(\mu_X)$, taken from 
Tab.~\ref{table:spectra}, for each $X$. 
Vertical lines represent the corresponding scales $\mu_X$ taken
from the same table. Also $\alpha_S^{(4)}(q)$ is shown by a dashed line.
\label{fx2}}
\end{figure}
The support function $f_X$ is given by
\bea
f_X(q) = \theta (\overline{m}-q) - \int_0^\infty dr \, r^2 | R_X(r) |^2 
\, \frac{\sin (qr)}{qr},
\eea
where $R_X(r)$ is the radial part of the Coulomb wave-function of $X$. 
$f_X(q)$ is almost unity in the region $1/a_X \simlt q < \overline{m}$, 
where $a_X$ is the inverse of the dumping scale of $f_X$ and may be
interpreted as the size of the boundstate $X$.
Thus, Eq.~(\ref{eth3}) can be identified with Eq.~(\ref{app-Etot})
qualitatively.

For the $1S$ state $f_X(q)$ dumps slowly 
as $q$ decreases. For other states $f_X(q)$ dumps rapidly from scales 
which are somewhat smaller 
than the naive expectations $(C_F \alpha_S m_{\rm pole})/n_X$.
In Fig.~\ref{fx2} we show $f_X$ for different states calculated with 
$m_{\rm pole}=5$ GeV 
and $\alpha_S(\mu_X)$ taken from Tab.~\ref{table:spectra}. 
The corresponding values 
of $\mu_X$ are also displayed. For those states which we consider the 
predictions reliable,
$\mu_X$ is located within the range where $f_X(q) \simeq 1$ 
(close to the infrared edge);
for those states with unreliable predictions, $\mu_X$ is located out of 
this range but far in the infrared region. 
This comparison shows a relation between the scale $\mu_X$ and
the wave function (consequently the size)
of the corresponding boundstate $X$.

\begin{figure}[t]
\begin{center}
    \includegraphics[width=12cm]{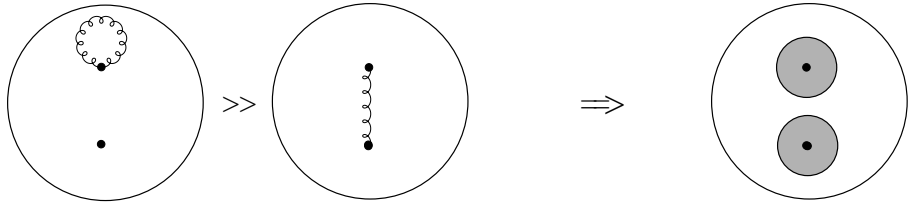}
\end{center}
\caption{\footnotesize
The total energy of a heavy quarkonium state is carried by the
$\overline{\rm MS}$ masses of quark and antiquark and by 
the gluons inside the boundstate.
In the latter contributions the self-energies of quark and antiquark 
dominate over the potential energy between the two particles.
\label{physpic}}
\end{figure}

Eq.~(\ref{app-Etot}) or (\ref{eth3}) tells that  {\it the major contribution 
to the energy levels comes from the region 
$1/a_X$ $ \simlt$ $ q$ $\simlt$ $\overline{m}$ of the self-energy corrections 
of quark and antiquark}
(apart from the constant contribution $2 \overline{m}$).
That is, with respect to only the 
contributions of the gluons inside the boundstate,
the self-energies of quark and antiquark dominate over the
potential energy between the two particles.
Hence, we find a qualitative picture on the composition of the energy
of a bottomonium state:
\begin{itemize}
\item[(I)]
The energy levels of bottomonium are mainly determined by 
(i) the $\overline{\rm MS}$ masses of $b$ and $\bar{b}$, and 
(ii) contributions to the self-energies of $b$ and $\bar{b}$
from gluons with wavelengths $1/\overline{m} \simlt \lambda \simlt a_X$.
The latter contributions may be regarded as the difference between
the (state-dependent) constituent quark masses and the current quark masses.
\end{itemize}
This picture is reminiscent of the (long-established)
interpretation that
the masses of light hadrons consist of the constituent quark masses.
See Fig.~\ref{physpic}.

Now we are in a position to see why the level spacings among the bottomonium
excited states are much wider than those of the Coulomb spectrum.
We know that if the coupling $\alpha_S(q)$
were independent of $q$, the energy spectrum would
become the Coulomb spectrum.
In Fig.~\ref{fx2} also $\alpha_S(q)$ is shown. 
We see that as $n_X$ 
increases from 1 to 3, the coupling $\alpha_S(q)$, close to the 
dumping scale of $f_X(q)$, grows rapidly.
According to Eq.~(\ref{app-Etot}) or (\ref{eth3}), as the integral region 
extends down to smaller $q$, the self-energy contributions grow
rapidly in comparison to the non-running case.
They push up the energy levels of the excited states considerably
and widen the level spacings among the excited states.
Thus, we may draw the following 
qualitative picture of the bottomonium level structure:
\begin{itemize}
\item[(II)]
Level spacing between consecutive  $n$'s increases rapidly with $n$
as compared to the Coulomb spectra. 
This is because the self-energy contributions (from
$1/\overline{m} \simlt \lambda \simlt a_X$)
grow rapidly as the physical size $a_X$ of the boundstate increases.
\end{itemize}

\section{A Link to Phenomenological Potentials}

We return to the problem which we discussed in the introduction (Sec.~1):
How can we understand the ``linear potential'' in the difference
between the Coulomb potential and a typical phenomenological
potential?

According to the interpretation given
in the previous section, the total energy 
$E_{\rm tot}(r) \equiv 2 m_{\rm pole} + V_{\rm QCD}(r)$
of a $b\bar{b}$ system determines the bulk of the bottomonium spectrum.
Hence, we examine the series expansion in $\alpha_S(\mu)$ of the total energy, 
expressed in terms of the $b$-quark $\overline{\rm MS}$ mass,
$E_{\rm tot}(r;\overline{m}_b,\alpha_S(\mu))$, and compare it
with phenomenological potentials \cite{connection}.
The obtained total energy depends on the scale $\mu$ due to 
truncation of the series at ${\cal O}(\alpha_S^3)$.
One finds that, when $r$ is small, the series
converges better and the value of $E_{\rm tot}(r)$ is less $\mu$-dependent
if we choose a large scale for $\mu$, whereas when $r$ is larger,
the series converges better and the value of $E_{\rm tot}(r)$ is 
less $\mu$-dependent if we choose a smaller scale for $\mu$.
Taking into account this property, 
we will fix the scale $\mu$ in two different ways below:
\begin{enumerate}
\item
We fix the scale $\mu = \mu_1(r)$ by demanding stability against variation of 
the scale:
\bea
\left. \mu \frac{d}{d\mu} E_{\rm tot}(r;\overline{m},\alpha_S(\mu))
\right|_{\mu = \mu_1(r)} = 0 .
\label{scalefix1}
\eea
\item
We fix the scale $\mu = \mu_2(r)$
on the minimum of the absolute value of the last known term 
[${\cal O}(\alpha_S^3)$ term] of $E_{\rm tot}(r)$:
\bea
\left. \mu \frac{d}{d\mu}
\Bigl[ E^{(3)}_{\rm tot}(r;\overline{m},\alpha_S(\mu)) \Bigr]^2 \,
\right|_{\mu = \mu_2(r)} = 0 .
\label{scalefix2}
\eea
\end{enumerate}
\begin{figure}[t]
\begin{center}
    \includegraphics[width=12cm]{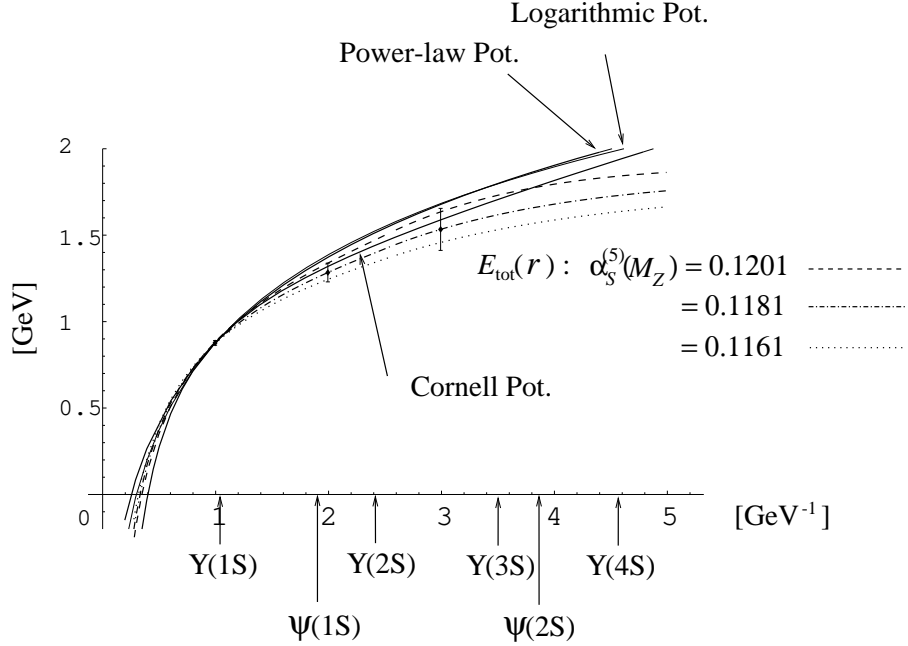}
\end{center}
\caption{\footnotesize
A comparison of the total energy of a $b\bar{b}$ system 
and typical phenomenological potentials (solid lines).
Effects of the non-zero charm quark mass are included.
For a reference, we show typical sizes of the bottomonium and charmonium
$S$ states as determined from the r.m.s.\ interquark distances with respect to
the Cornell potential: $\sqrt{\left< r^2 \right>}_{\rm Cornell}$.
\label{comppheno-mceff}
}
\end{figure}
It turns out that 
the total energy hardly changes whether we choose $\mu=\mu_1(r)$
or $\mu=\mu_2(r)$;
also the scales $\mu_1(r)$ and $\mu_2(r)$ are considerably larger than $1/r$;
the perturbative prediction for $E_{\rm tot}(r)$ becomes unstable
at $r \simgt 3$~GeV$^{-1}$.

In Fig.~\ref{comppheno-mceff} we compare the total energy 
for $\alpha_S^{(5)}(M_Z)=0.1201$, 0.1181, 0.1161 with
typical phenomenological potentials:
\begin{itemize}
\item
A Coulomb-plus-linear potential (Cornell potential) \cite{cornell}:
\bea
V(r) = - \frac{\kappa}{r} + \frac{r}{a^2}
\eea
with $\kappa = 0.52$ and $a = 2.34$~GeV$^{-1}$.
\item
A power-law potential \cite{martin}:
\bea
V(r) = - 8.064~{\rm GeV} +
(6.898~{\rm GeV})(r\times 1~{\rm GeV})^{0.1} .
\eea
\item
A logarithmic potential \cite{qr}:
\bea
V(r) = -0.6635~{\rm GeV} + (0.733~{\rm GeV}) \log (r\times 1~{\rm GeV}) .
\eea
\end{itemize}
In order to make a clear comparison, arbitrary constants have been added
to all the potentials and $E_{\rm tot}(r)$
such that their values coincide at $r=1$~GeV$^{-1}$.
The total energy appears to be in good agreement with the phenomenological
potentials in the range
$0.5~{\rm GeV}^{-1} \simlt r \simlt 3~{\rm GeV}^{-1}$.
The level of agreement is consistent with
the uncertainties expected from
the next-to-leading renormalon contributions 
$\Bigl[$$\pm {\frac{1}{2}} \Lambda_{\rm QCD} \cdot ( \Lambda_{\rm QCD} r )^2$ 
taking $\Lambda_{\rm QCD} = 300$~MeV: indicated by the error bars$\Bigr]$.

Thus, we confirm that, when the renormalon cancellation is incorporated, 
the QCD radiative corrections bend
the Coulomb potential upwards at long distances.
One may understand it similarly to the previous discussion:
When the infrared cutoff $\sim 1/r$ of 
Eqs.~(\ref{total-ene})--(\ref{renormalon-mass}) is lowered,
i.e.\ at large $r$,
$\alpha_S(q)$ grows rapidly in the integrand,
which raises the total energy as compared to the non-running (Coulomb) case.

Another way to understand it qualitatively
is to consider the interquark force defined by
\bea
F(r) &\equiv& 
- \frac{d}{dr} E_{\rm tot}(r) = - \frac{d}{dr} V_{\rm QCD}(r)
\\
& \equiv &
- C_F \frac{\alpha_F (1/r)}{r^2} . 
\eea
The last line defines the ``$F$-scheme'' coupling constant $\alpha_F(\mu)$.
The interquark force is also free from the leading renormalon.
The running of $\alpha_F(\mu)$ is dictated by 
the renormalization-group equation:
\bea
\mu^2 \frac{d}{d\mu^2} \alpha_F (\mu) = \beta_F (\alpha_F) ,
\label{betaF}
\eea
where the first two coefficients of the beta function are universal,
i.e.\ same as those of $\beta_{\overline{\rm MS}}(\alpha_S)$.
When we consider effects of the QCD radiative corrections on the 
lowest-order Coulomb potential,
one may interpret that in the QCD potential,
$V_{\rm QCD}(r) \simeq -C_F \alpha_S(1/r)/r$, the coupling increases at long
distances, so the potential will be bent downwards.
This is a bad interpretation, since the QCD potential is poorly
convergent at relevant distances.
We should rather consider the interquark force.
A better interpretation is that in 
$F(r)=-C_F \alpha_F(1/r)/r^2$, the $F$-scheme coupling increases at long
distances, and correspondingly $|F(r)|$ grows with respect to the
Coulomb force.
This means that the slope of the potential becomes steeper at
long distances.
Its effect resembles an addition of a linearly rising potential
to the Coulomb potential.
Thus, the effects of the radiative corrections are
even qualitatively reversed, whether we consider
$V_{\rm QCD}(r)$ or $F(r)$ as the physically relevant quantity.

\begin{figure}[tbp]
  \hspace*{\fill}
    \includegraphics[width=10cm]{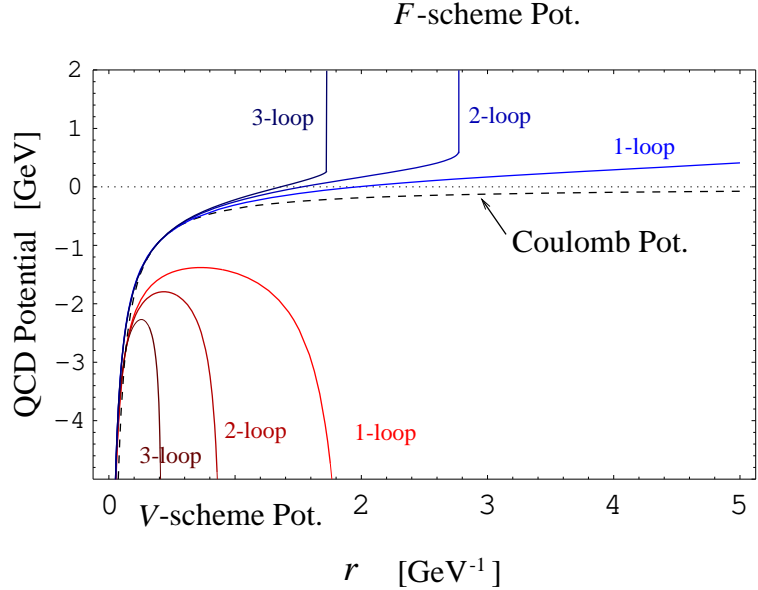}
  \hspace*{\fill}
  \\
  \hspace*{\fill}
\caption{\footnotesize
A comparison of the QCD potentials calculated in $V$-scheme and in
$F$-scheme as well as the Coulomb potential.
We set $m_c=0$ and consider 4 massless flavors.
The Coulomb potential is given by $-C_F \alpha /r$ with
$\alpha = 0.279$.
      \label{fig-Fscheme}
}
  \hspace*{\fill}
\end{figure}
One may verify these features in Fig.~\ref{fig-Fscheme}, 
in which the Coulomb potential,
the $V$-scheme potentials and the $F$-scheme potentials are displayed.
The $V$-scheme potentials are calculated by solving the 
renormalization-group equation for the QCD potential.
The $F$-scheme potentials are calculated by first solving the 
renormalization-group equation (\ref{betaF})
for $\alpha_F$ numerically
and then by integrating $-F(r)$ over $r$ numerically;
arbitrary constants are added such that the $F$-scheme potentials 
coincide the Coulomb potential at $r=0.4$~GeV$^{-1}$.
As can be seen, the $V$-scheme potentials become singular at fairly
short distances, whereas 
the $F$-scheme potentials have wider ranges of validity.
The 2-loop and 3-loop $F$-scheme potentials are consistent with
the phenomenological potentials within the uncertainty
expected from the next-to-leading renormalon contributions,
in the range $0.5~{\rm GeV}^{-1} \simlt r \simlt 2.8~{\rm GeV}^{-1}$
and $0.5~{\rm GeV}^{-1} \simlt r \simlt 1.7~{\rm GeV}^{-1}$,
respectively.
On the other hand, the 1-loop $F$-scheme potential does not
satisfy this criterion.

\section{Conclusions and Discussion}

For all the bottomonium states, where the predictions of perturbative QCD 
can be made reliably (i.e. $\alpha_S < 1$), our results are consistent with 
the experimental 
data within the estimated uncertainties of the theoretical predictions. 
The obtained value 
for $\overline{m}_b$ is in good agreement with the recent sum-rule 
calculations.
The theoretical uncertainties given in Tab.~\ref{error} are numerically 
of the same size as $\Lambda_{\rm QCD}\times (a_X \Lambda_{\rm QCD})^2$,
i.e. of the effect of the next-to-leading renormalons: 
if we approximate $1/a_X \simeq \mu_X$, take the values of 
Tab.~\ref{table:spectra}, and $\Lambda_{\rm QCD} = 300 - 500$ MeV, we obtain 
for the $1S$ state a contribution of order $\pm(5 - 20)$ MeV, for the $n=2$ 
states a contribution 
of order $\pm(20 - 110)$ MeV and for the $3S$ state a contribution of order 
$\pm(50 - 250)$ MeV. 
Since the mass $\overline{m}_b$ has been fixed on the vector ground state 
and has {\it not} 
been adjusted for higher states, the data at our disposal suggest that:
1) the bulk of the bottomonium spectrum is accessible by perturbative 
QCD up to some of the $n=3$ states; 
2) non-perturbative contributions do not need to 
be larger than $250$ MeV 
for the reliable $n=3$ states, than $100$ MeV for the $n=2$ states and 
than $10$ MeV for $\overline{m}_b$, 
and they are consistent with 
the type associated with the next-to-leading renormalon. 
These upper bounds to the non-perturbative corrections are conservative 
and their true sizes may be considerably smaller; 
note that for reliable predictions all of $|E_X^{exp} - E_X|$ 
in Tab.~\ref{table:spectra} are smaller than 60 MeV. 

When we incorporate the cancellation of the leading renormalon
contributions,
the perturbative expansion of the total energy $E_{\rm tot}(r)$
of a $b\bar{b}$ system, up to ${\cal O}(\alpha_S^3)$ and supplemented
by the scale-fixing prescription (\ref{scalefix1}) or (\ref{scalefix2}), 
converges well at $r \simlt 3~{\rm GeV}^{-1}$.
Moreover, it agrees with the phenomenologically determined potentials
in the range 
$0.5~{\rm GeV}^{-1} \simlt r \simlt 3~{\rm GeV}^{-1}$
within the uncertainty expected from the next-to-leading renormalon
contributions.
This establishes a connection between our present approach based on
perturbative QCD and the conventional phenomenological potential-model
approaches.
Eventually we may merge them
and further develop understandings of the charmonium and bottomonium
systems.
For instance, in the perturbative prediction for the bottomonium
spectrum, the level splittings 
between the $S$-wave and $P$-wave states as well as
the fine splittings among the $nP_j$ states are
smaller than the corresponding experimental values.
Although the discrepancy is still smaller than the estimated theoretical
uncertainties of the predictions, it should certainly be
clarified whether they are explained by higher-order perturbative
corrections, or, we need specific non-perturbative effects for
describing them.
On the other hand, the conventional potential-model approaches have been
successful also in explaining the $S$-$P$ splittings and the fine splittings.
Hence, we expect that the connection would help 
to clarify origins of the differences of the present
perturbative predictions and the experimental data.

For what concerns the $c\bar{c}$ system, we obtain 
$\overline{m}_c = 1243 \pm 15 \pm 20 \pm 50~{\rm MeV}$
from the mass of the $J/\psi$ state:
the first error is due to the uncertainty in $\alpha_S(M_Z)$,
the second error is due to higher-order corrections,
and the third error is due to non-perturbative contributions.
We note that this estimate is in good agreement with 
recent sum-rule calculations. 
With the present method, however, we cannot
make reliable predictions for states higher than the ground state
of charmonium, and, therefore, 
we cannot extrapolate from consistency arguments the size of the 
non-perturbative corrections. 
Nevertheless, from the prediction of the $\eta_c$ mass, 
we may anticipate non-perturbative contributions of the order of $100$ MeV
to the $n=1$ states. 
Also this figure is consistent 
with the next-to-leading renormalon effect ($20 - 110$ MeV).
Our prediction for the mass of the $B_c(1^1S_0)$ state is
$M_{B_c(1S)}=6324 \pm 5 \pm 20 \pm 40$~MeV.

We provided a novel picture on the composition of the masses
of the bottomonium states.
The picture may be contrasted to that of the QED boundstates such
as positronium.
For this system, the pole masses of a free electron and positron
are well-defined.
Then the boundstate mass is given by the sum of the pole masses
minus the binding energy.
For a bottomonium state, we cannot define the pole masses by
separating $b$ and $\bar{b}$ an infinite distance.\footnote{
The linear rise at $r \simgt 1~{\rm GeV}^{-1}$ 
and an instability at $r \simgt 3~{\rm GeV}^{-1}$
of the perturbative prediction for $E_{\rm tot}(r)$ may be
taken as indicative signatures of quark confinement,
at the best of the state of the art of the perturbative QCD prediction.
}
Therefore, the only sensible description is through contributions of the
gluons which reside inside the boundstate.
In this picture: the mass of the boundstate becomes 
heavier than the sum of
the quark $\overline{\rm MS}$ masses ($\approx$ current quark masses);
it is composed of the sum of the $\overline{\rm MS}$ masses
and the self-energies ($\approx$ constituent quark masses), 
supplemented by the small negative potential energy which
binds the system together.
It would be an interesting question if this picture is also
applicable to lighter QCD boundstates.

\section*{Acknowledgements}
Most of the results reported here are based on the
collaboration with N.~Brambilla and A.~Vairo.
The author is grateful for very fruitful discussions.

\def\app#1#2#3{{\it Acta~Phys.~Polonica~}{\bf B #1} (#2) #3}
\def\apa#1#2#3{{\it Acta Physica Austriaca~}{\bf#1} (#2) #3}
\def\npb#1#2#3{{\it Nucl.~Phys.~}{\bf B #1} (#2) #3}
\def\plb#1#2#3{{\it Phys.~Lett.~}{\bf B #1} (#2) #3}
\def\prd#1#2#3{{\it Phys.~Rev.~}{\bf D #1} (#2) #3}
\def\pR#1#2#3{{\it Phys.~Rev.~}{\bf #1} (#2) #3}
\def\prl#1#2#3{{\it Phys.~Rev.~Lett.~}{\bf #1} (#2) #3}
\def\sovnp#1#2#3{{\it Sov.~J.~Nucl.~Phys.~}{\bf #1} (#2) #3}
\def\yadfiz#1#2#3{{\it Yad.~Fiz.~}{\bf #1} (#2) #3}
\def\jetp#1#2#3{{\it JETP~Lett.~}{\bf #1} (#2) #3}
\def\zpc#1#2#3{{\it Z.~Phys.~}{\bf C #1} (#2) #3}

\end{document}